\documentclass[10pt]{article}
\usepackage[dvips]{epsfig}
\usepackage{epsf}
\newcommand{\beq}{\begin{equation}}

\newcommand{\eeq}{\end{equation}}
\newcommand{\bea}{\begin{eqnarray}}
\newcommand{\eea}{\end{eqnarray}}
\begin{document}

\begin{center}

{\large \bf
Analysis of the GSI A+p and A+A Spallation, Fission, and 
Fragmentation Measurements with the LANL CEM2k and LAQGSM codes
}\\

\vspace{0.5cm}

S.G.~Mashnik$^{1,}$\footnote[3]{mashnik@lanl.gov},
K.K.~Gudima$^{2}$, 
R.E. Prael$^1$,  A.J.~Sierk$^1$

\vspace{0.2cm}
$^1${\em 
Los Alamos National Laboratory, Los Alamos, NM 87545, USA}\\

$^2${\em Institute of Applied Physics,
Academy of Science of Moldova, Kishinev, MD-2028, Moldova}
\end{center}

\vspace{0.3cm}
%\begin{center}
%{\bf Abstract}
%\end{center}
{\small \noindent
{\bf Abstract:}
The CEM2k and LAQGSM codes have been recently developed at
Los Alamos National Laboratory to simulate nuclear reactions 
induced by particles and nuclei for a
number of applications. We have benchmarked our codes
against most available measured data at projectile energies 
from 10 MeV/A to 800 GeV/A and have compared our results with predictions
of other current models used by the nuclear community. Here, we present
a brief description of our codes and show illustrative results obtained
with CEM2k and LAQGSM for A+p and A+A spallation, fission, and 
fragmentation reactions measured recently at GSI compared with 
predictions by other models. Further necessary work is outlined.}

\vspace*{5mm}
{\noindent \bf \large Introduction}\\

During recent years, for a number of applications like 
Accelerator Transmutation of nuclear Waste (ATW),
Accelerator Production of Tritium (APT),
Rare Isotope Accelerator (RIA), Proton Radiography (Prad),
astrophysical work for NASA, and other projects,
we have developed at the Los Alamos National Laboratory
an improved version of the Cascade-Exciton Model (CEM), 
contained in the code CEM2k, to describe  nucleon-, pion-, 
and photon-induced reactions at incident energies up to 
about 5 GeV \cite{CEM2k}-\cite{fitaf}
and the Los Alamos version of the Quark-Gluon String Model,
realized in the high-energy code LAQGSM \cite{LAQGSM},
to describe both particle- and nucleus-induced reactions
at energies up to about 1 TeV/nucleon \cite{LAQGSM}-\cite{Mashnik02d}.

Both codes have been tested against most of the available data
and compared with predictions of other modern codes
\cite{CEM2k}-\cite{Mashnik02d}. 
Our comparisons show that these codes describe 
a large variety of spallation, fission, and fragmentation reactions
quite well and often have a better predictive power than some
other available Monte-Carlo codes, thus they can be used
as reliable event generators in different applications and in
fundamental nuclear research.

We have analyzed with CEM2k and LAQGSM 
all the A+p and A+A measurements done recently at GSI 
at energies near or below 1 GeV/nucleon for which we
have results.
The size of this paper allows us to present only a brief description of 
our models and a few results for proton-nucleus and  nucleus-nucleus 
spallation, fission and fragmentation reactions measured at GSI. 
Results for other reactions may be found in Refs. 
\cite{CEM2k}-\cite{Mashnik02d}. \\

{\noindent \bf \large CEM2k and LAQGSM Codes}\\

A detailed description of the initial version of the CEM may be found
in \cite{Gudima83}, therefore we outline here only its basic
assumptions.
The CEM assumes that reactions occur in three stages. The first
stage is the IntraNuclear Cascade (INC) 
in which primary particles can be re-scattered and produce secondary
particles several times prior to absorption by or escape from the nucleus.
The excited residual nucleus remaining after the 
cascade determines the particle-hole configuration that is
the starting point for the preequilibrium stage of the
reaction. The subsequent relaxation of the nuclear excitation is
treated in terms of an improved Modified Exciton Model (MEM) of preequilibrium 
decay followed by the equilibrium evaporative final stage of the reaction.
Generally, all three stages contribute to experimentally measured outcomes.

The improved cascade-exciton model in the code CEM2k differs from the 
older CEM95 version \cite{CEM95} by incorporating new 
approximations for the elementary cross sections used in the cascade,
more precise values for nuclear masses and pairing energies, 
a corrected systematics for the level-density parameters, 
adjusted cross sections for pion absorption on quasi-deuteron 
pairs inside a nucleus, 
the Pauli principle in the preequilibrium calculation, 
and an improved calculation of fission widths.
Significant refinements and improvements in the 
algorithms used in many subroutines lead to a decrease of
computing time by up to a factor of 6 for heavy nuclei, which 
is very important when performing simulations with transport codes.
Essentially, CEM2k has a longer cascade stage,
less preequilibrium emission, and a longer evaporation stage
with a higher initial excitation energy, compared to its precursors
CEM97 \cite{CEM97} and CEM95 \cite{CEM95}.
Besides the changes to CEM97 and CEM95 mentioned, we also made a 
number of other improvements and refinements, such as:
(i) imposing momentum-energy conservation for each simulated 
event (the Monte-Carlo algorithm previously used in CEM 
provided momentum-energy conservation only 
statistically, but not exactly for the cascade stage 
of each event),
(ii) using real binding energies for nucleons at the cascade 
stage instead of the approximation of a constant
separation energy of 7 MeV used in previous versions of the CEM,
(iii) using reduced masses of particles in the calculation of their
emission widths instead of using the approximation
of no recoil used previously, and
(iv) a better approximation of the total reaction cross sections.
On the whole, this set of improvements leads to a much better description
of particle spectra and yields of residual nuclei and a better 
agreement with available data for a variety of reactions.
Details, examples, and further references may be found in
\cite{CEM2k}-\cite{fitaf}.

The Los Alamos version of the Quark-Gluon String Model
(LAQGSM) \cite{LAQGSM} is a further development of 
the Quark-Gluon String Model (QGSM) by Amelin, Gudima, and Toneev
(see \cite{Amelin90} and references therein) and is intended to describe
both particle- and nucleus-induced reactions at energies up to
about 1 TeV/nucleon. 
The core of the QGSM is built on a time-dependent version of the
intranuclear-cascade model developed at Dubna,
often referred in the literature simply
as the Dubna intranuclear Cascade Model (DCM) (see \cite{Toneev83}
and references therein).
The DCM models interactions of fast cascade particles (``participants")
with nucleon spectators of both the target and projectile nuclei and
includes interactions of two participants (cascade particles) as well.
It uses experimental cross sections (or those calculated by the Quark-Gluon 
String Model for energies above 4.5 GeV/nucleon) for these
elementary interactions to simulate angular and energy distributions
of cascade particles, also considering the Pauli exclusion principle. 
When the cascade stage of a reaction is completed, QGSM uses the
coalescence model described in \cite{Toneev83}
to ``create" high-energy d, t, $^3$He, and $^4$He by
final-state interactions among emitted cascade nucleons outside 
of the colliding nuclei.
After calculating the coalescence stage of a reaction, QGSM
moves to the description of the last slow stages of the reaction,
namely to preequilibrium decay and evaporation, with a possible competition
of fission using the standard version of the CEM \cite{Gudima83}.
If the residual nuclei have atomic numbers 
with  $A \le 13$, QGSM uses the Fermi break-up model 
to calculate their further disintegration instead of using
the preequilibrium and evaporation models.
LAQGSM differs from QGSM by replacing the preequilibrium and
evaporation parts  of QGSM described according to the standard CEM 
\cite{Gudima83} with the new physics from CEM2k
\cite{CEM2k,CEM2kTsukuba} and has a number of improvements 
and refinements in the cascade, coalescence,
and the Fermi break-up models (in the
current version of LAQGSM, we use the Fermi break-up model only for
 $A \le 12$). A detailed description of LAQGSM and further
references may be found in \cite{LAQGSM} and in our later publications
\cite{Mashnik02a,fitaf,OurNewINC}.

Originally, both CEM2k and LAQGSM were not able to describe fission reactions 
and production of light fragments heavier than $^4$He, as they had neither 
a high-energy-fission nor a fragmentation model.  Recently, we addressed 
these problems \cite{Mashnik02a}-\cite{fitaf} by further improving
our codes and by merging them with the Generalized Evaporation Model
code GEM2 developed by Furihata \cite{GEM2}.

Our current versions of CEM2k and LAQGSM were
incorporated recently into the MARS 
%\cite{MARS} 
and LAHET 
%\cite{LAHET}
transport codes and are currently being incorporated into MCNPX.
% \cite{MCNPX}.
This will allow others to use our codes as event-generators
in these transport codes to simulate reactions with targets of
practically arbitrary geometry and nuclide composition.\\

{\noindent \bf \large Illustrative Results}\\

The Generalized Evaporation Model
code GEM2 of Furihata \cite{GEM2} merged with both our
CEM2k and LAQGSM takes into account evaporation
of up to 66 types of  particles and light fragments (from
n to $^{28}$Mg) from excited compound nuclei, while most other
evaporation models used in the literature consider evaporation
of only 6 types of particles, from n to $^4$He. It is interesting to
see how important this is when analyzing the recent GSI 
A+p and A+A measurements.
Fig. 1 gives us a quick and clear answer to this question: {\it It is not 
important at all}.
We see that
calculations by CEM2k+GEM2 taking into account up to 
66 types of evaporated particles and fragments almost coincide
with similar results calculated considering only 6 types of
evaporated particles for all products
measured recently at GSI \cite{Rejmund01} for the reaction
800 MeV/A $^{197}$Au+p. Similar results were obtained for other
GSI measurements. We do see a big difference between the
results of these ``66" and ``6" calculations, but only for 
products with $A < 28$ and $Z < 12$, and such light products
have not been measured at GSI. Note that the ``66"
calculations require about 7-8 times more computing time than
the ``6" ones. This means that if we study only spallation
and fission products and are not interested in light fragments, we
can consider evaporation of only 6 types of particles and save
the computing time, getting results very close
to the ones calculated with the more time consuming ``66" option.
But if we need to describe correctly all products from a reaction, including
the light fragments, we will need to use the ``66" option. 
All results presented below (and in previous 
publications) were obtained using the ``66" option.

Fig. 2 shows an example of a proton-induced reaction calculated
by LAQGSM+GEM2 measured at GSI in inverse kinematics, i.e., as
1 GeV/A $^{208}$Pb+p \cite{Enqvist01}. For comparison, results
obtained with the transport code LAHET3 \cite{LAHET3}
using the Liege 
intranuclear cascade (INC) model INCL by Cugnon {\it et al.} \cite{INCL}
merged with the GSI

%%%%%%%%%%%%%%%%%%%%%%%%%%%%%%%%%%%%%%%%%%%%%%%%%%%%%%%%%%%%%% End of p. 2

%\newpage
%**************************** Begin Figs. 1,2 **************************
\begin{figure}[h]
\begin{minipage}{11.0cm}
\vbox to 158mm {
\vspace*{-25mm}
\hspace*{-10mm}
\includegraphics[width=120mm,angle=-0]{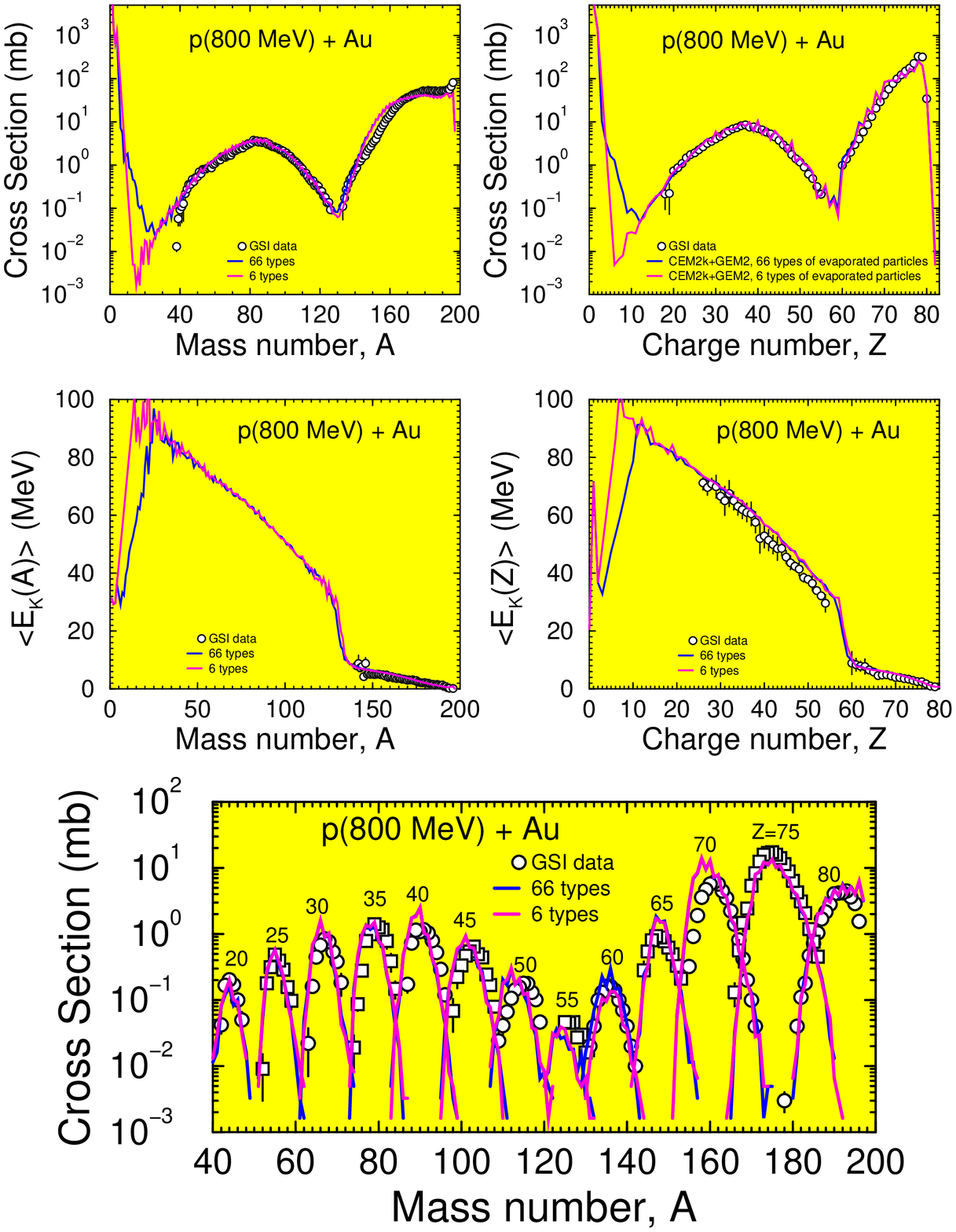}}
\end{minipage}
\hfill
\begin{minipage}{6.7cm}
\vspace*{-15mm}
\begin{small}
{\bf Figure 1.}
The measured \cite{Rejmund01}
mass 
and charge distributions of the product yields from the
reaction 800 MeV/A $^{197}$Au+p and of 
the mean kinetic energy of these products, and 
the mass distributions of the cross sections for 
the production of 
thirteen isotopes with the charge Z from 20 to 80 (circles)
compared with CEM2k+GEM2 calculations taking into account
evaporation up to 66 types of particles and light fragments
(blue lines) and considering evaporation of only 
n, p, d, t, $^3$He, and $^4$He (red lines).\\
\\
\end{small}

\vspace*{-5mm}

\hspace*{-1.8mm}
evaporation/fission model ABLA by
Schmidt {\it et al.} \cite{ABLA}, as incorporated into LAHET3 by
J.-C. David of CEA, Saclay (see some details in Appendix 2).

\hspace*{3mm}
Comparison of our CEM2k+GEM2 and LAQGSM+GEM2 results with
the third and the last proton-induced reaction measured at GSI
in inverse kinematics we found tabulated data, namely
1 GeV/A $^{238}$U+p \cite{Taieb02,Bernas03} 
is shown in Figs. A1.1-A1.4 of Appendix 1.
For convenience, we divide the results for the products
shown in these figures into three groups: spallation,
fission, and fragmentation.
We compare our results with  
calculation by LAHET3 \cite{LAHET3}
using the INCL+ABLA \cite{INCL,ABLA}
option mentioned above, as well as using the
ISABEL \cite{ISABEL} intranuclear  cascade model coupled with the
Dresner evaporation code \cite{Dresner} and 
the RAL fission model of Atchison \cite{RAL},
as well as using the Bertini INC \cite{Bertini}
coupled with Dresner \cite{Dresner} and  Atchison \cite{RAL}
models.
\end{minipage}

\begin{minipage}{7.0cm}
\vbox to 7.0cm {
\vspace*{-30mm}
\hspace*{-10mm}
\includegraphics[width=90mm,angle=-0]{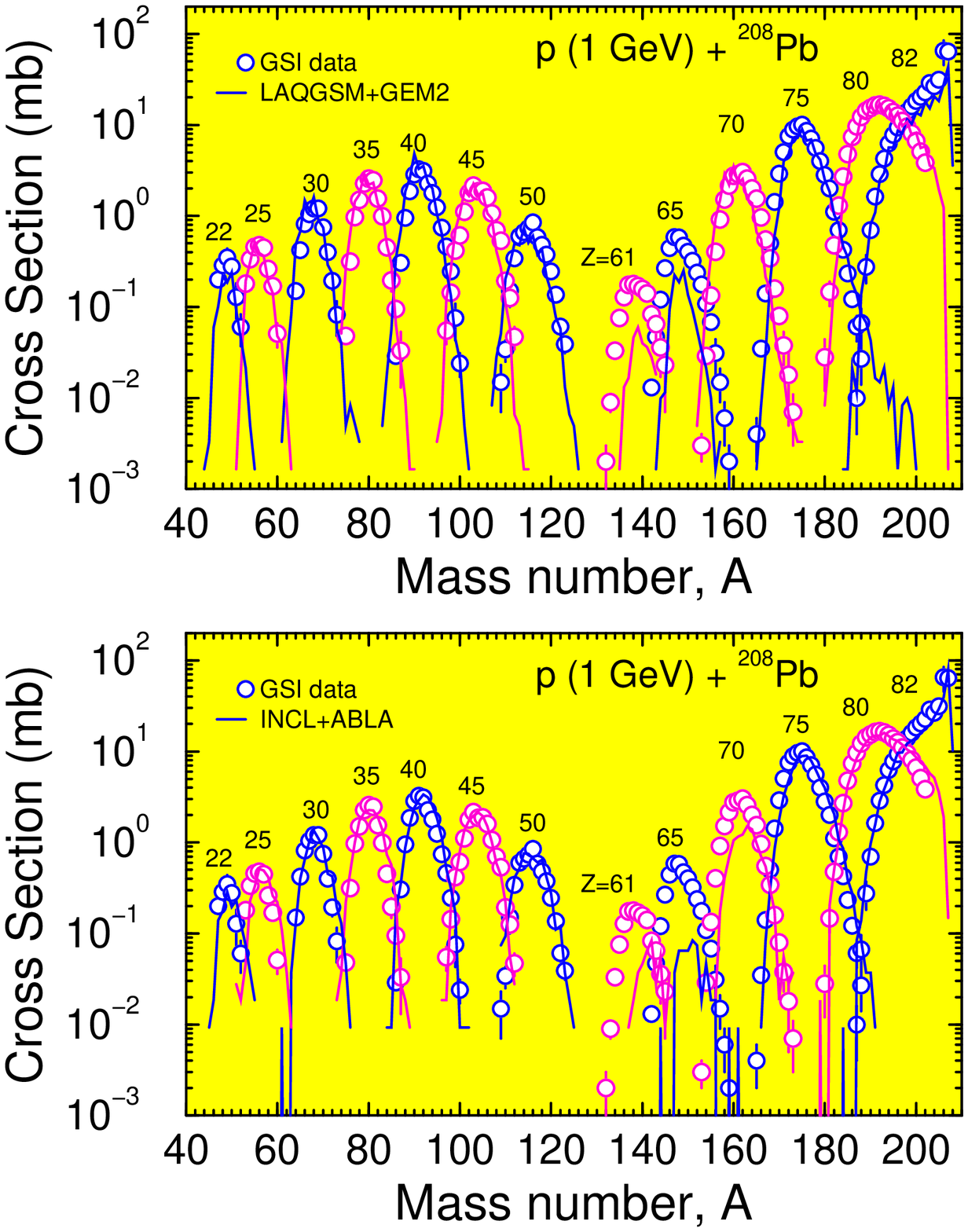}}
\end{minipage}
\hfill
\begin{minipage}{9.5cm}
\vspace*{-5mm}
\begin{small}
{\bf Figure 2.}
Experimental \cite{Enqvist01} mass distributions of the cross sections 
of thirteen isotopes with the charge $Z$ from 22 to 82
compared with our LAQGSM+GEM2 calculations
and results by LAHET3 \cite{LAHET3} using the Cugnon {\it et al.}
intranuclear cascade model INCL \cite{INCL}
merged with the GSI evaporation/fission model ABLA by
Schmidt {\it et al.} \cite{ABLA}.
\\
\\
\end{small}

\vspace*{-5mm}
\hspace*{3mm}
We performed all calculations of this reaction in 2002
(except with INCL+ABLA, see details in Appendix 2), after the 
measured spallation product cross sections
were published in \cite{Taieb02}, and published part of these results in
a 2002 LANL Theoretical Division Report of Activity \cite{T&NW2002}.
The experimental data on fission and fragmentation products were 
published only in 2003 \cite{Bernas03};
therefore the CEM2k+GEM2, LAQGSM+GEM2,
ISABEL+Dresner/Atchison, and Bertini+Dresner/Atchison 
results for fission and fragmentation products shown 
in  Figs. A1.1-A1.4 are pure predictions, many of which
agree amazingly well with the experimental data.

\end{minipage}

\end{figure}
%**************************** End Figs. 1,2  *************************

%\newpage
%**************************** Begin Figs. 3,4 **************************
\begin{figure}[h]
\begin{minipage}{7.0cm}
\vbox to 12.0cm {
\vspace*{-20mm}
\hspace*{-23mm}
\includegraphics[width=100mm,angle=-0]{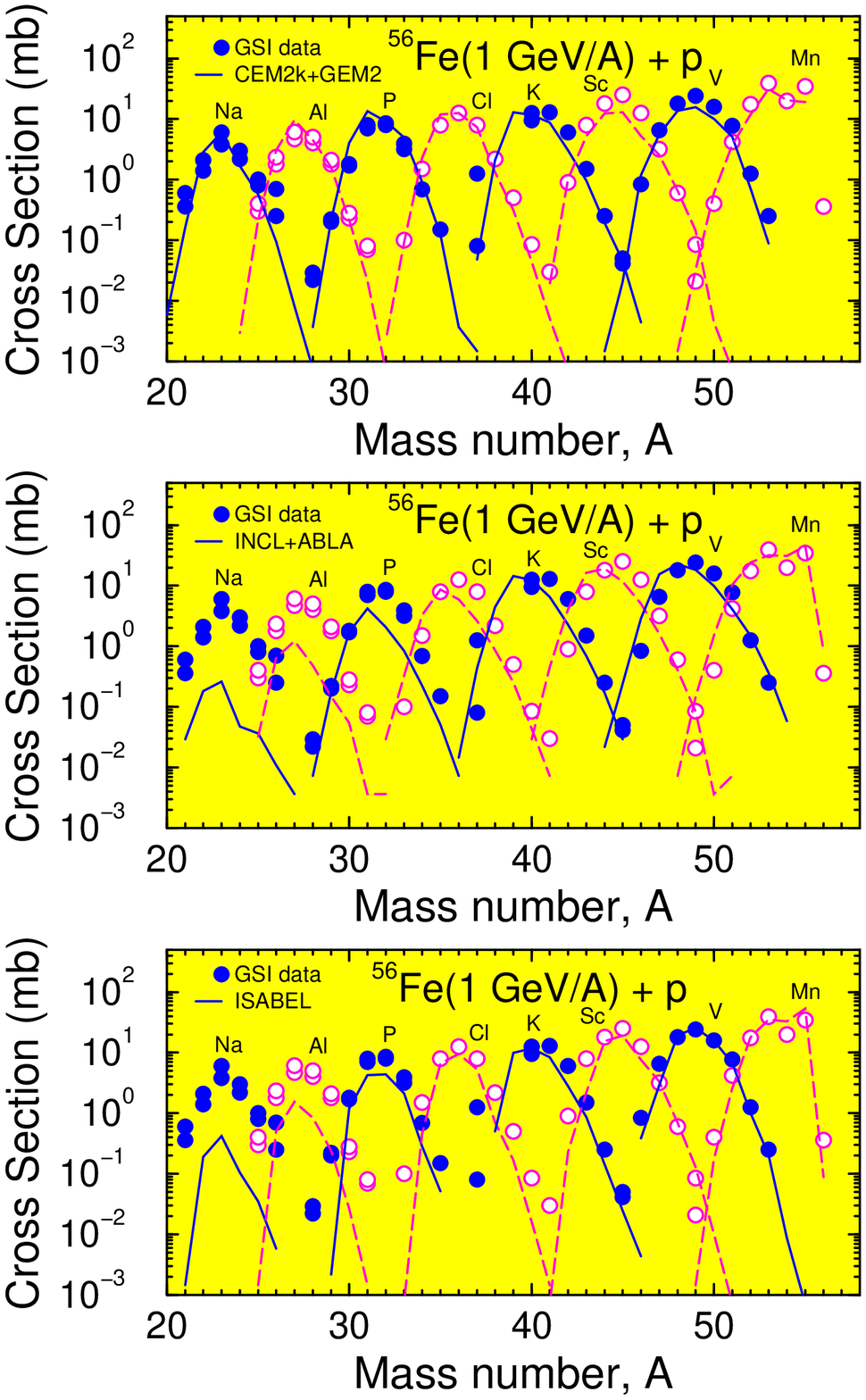}}
\end{minipage}
\hfill
\begin{minipage}{10.5cm}
\vspace*{-28mm}
\begin{small}
{\bf Figure 3.}
Comparison of preliminary experimental data
taken by us from Fig. 5 of Carmen Villagrassa
\cite{Carmen}
on mass distribution of the yields of eight isotopes
from Na to Mn produced in the reaction 1 GeV/A $^{56}$Fe + p (circles)
with our CEM2k+GEM2 results and with calculations by LAHET3
using the INCL+ABLA and ISABEL+Dresner/Atchison options (lines),
respectively.\\
\\
\end{small}
\\

\vspace*{-7mm}
\hspace*{3mm}
As do all other models, CEM2k+GEM2 and LAQGSM+GEM2
have many parameters. But all these parameters are fixed
and all the results shown in the figures of this paper were 
calculated within a single approach, without fitting any parameters:
We changed only the values of the mass and charge numbers of the
projectile and target nuclei and the incident energy
of the projectile in the input files of our codes.

\hspace*{3mm}
Fig. 3 shows the last proton-induced reaction we discuss in
the present work, 1 GeV/A $^{56}$Fe+p. In a way, our 
calculations shown in this figure can be considered  also as
predictions, as we do not have numerical values for 
the GSI measurements of this reaction and the circles shown 
in Fig. 3 as ``GSI data" are taken by us from 
Fig. 5 of Carmen Villagrassa \cite{Carmen} and are only preliminary.

\hspace*{3mm}
In the rest of the paper
we focus at nucleus-nucleus reactions measured recently
at GSI, and we start our analysis with the lightest target, d, namely
with the reaction $^{208}$Pb(1 GeV/A) + d  \cite{Enqvist02}
shown in Fig. 4.

\end{minipage}

\begin{minipage}{7.0cm}
\vbox to 12.0cm {
\vspace*{-15mm}
\hspace*{-23mm}
\includegraphics[width=100mm,angle=-0]{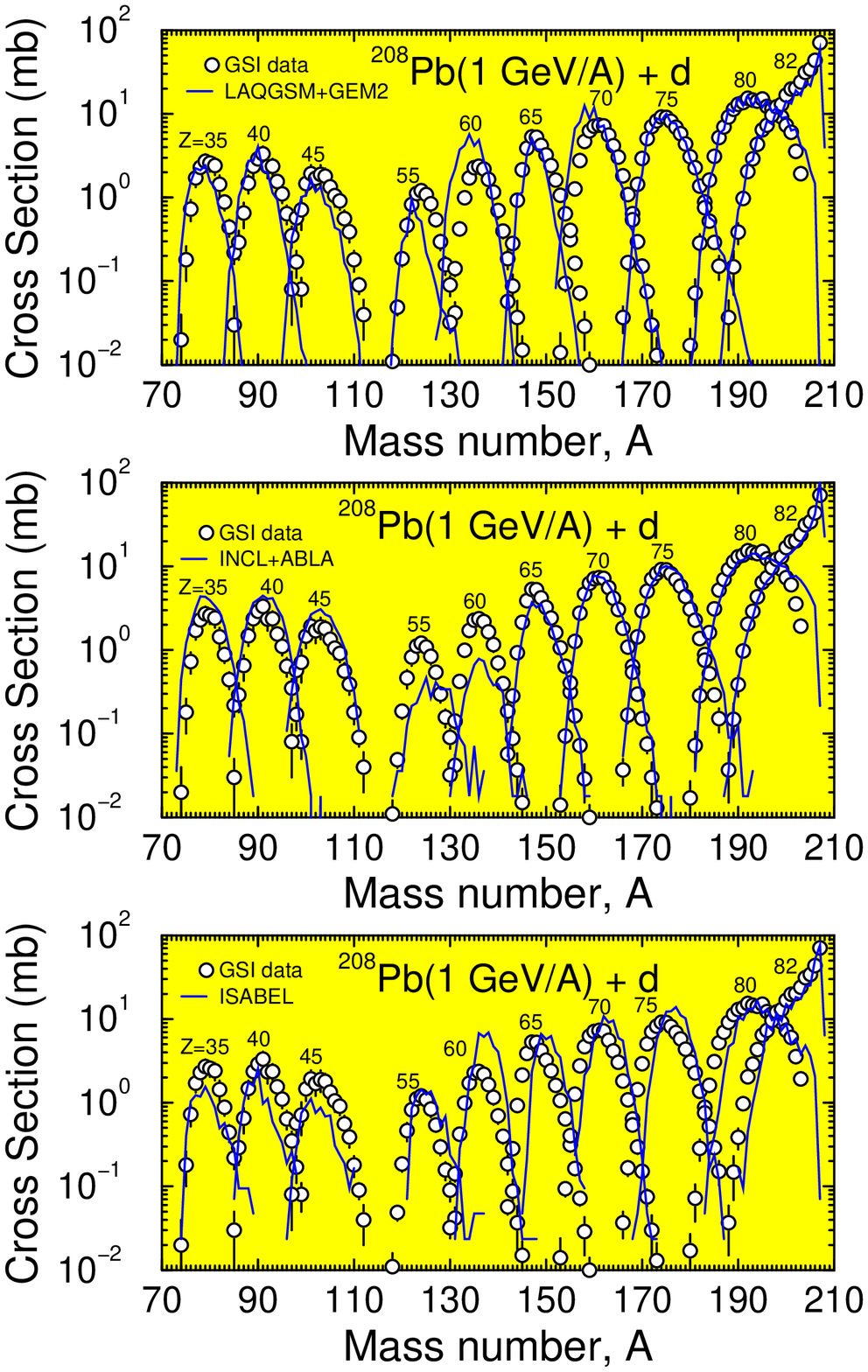}}
\end{minipage}
\hfill
\begin{minipage}{10.5cm}
\vspace*{-15mm}
\begin{small}
{\bf Figure 4.}
Experimental \cite{Enqvist02} mass distributions of the cross sections 
of ten isotopes with the charge $Z$ from 35 to 82 (circles)
produced in the reaction 1 GeV/A $^{208}$Pb + d
compared with our LAQGSM+GEM2
results and with calculations by LAHET3
using the INCL+ABLA and ISABEL+Dresner/Atchison options (lines),
respectively.\\
\\
\end{small}

\vspace*{-4mm}
One can see that LAQGSM+GEM2
describes quite well both the spallation and fission product
cross sections and agrees with most
of the GSI data with an accuracy of a factor of two or better
similarly to the results of the INCL+ABLA and ISABEL+Dresner/Atchison 
models.

\hspace*{3mm}
Fig. A1.5 in Appendix 1 shows another reaction on d: 1 GeV/A $^{238}$U+d.
Only the yields of spallation products from this reaction measured at GSI
are available to us \cite{Casarejos}. In Fig. A1.5, we compare our
LAQGSM+GEM2 results with all published data and   
with calculations by LAHET3
using the INCL+ABLA and ISABEL+Dresner/Atchison options.
We see that as in the case of Pb+d, LAQGSM+GEM2 describes the U+d data 
quite well, better than do the the INCL+ABLA and 
ISABEL+Dresner/Atchison models.

\hspace*{3mm}
Our CEM2k+GEM2 and LAQGSM+GEM2 codes describe correctly not only
the cross sections of the spallation, fission, and fragmentation
products from various reactions, but also their mean kinetic 
(recoil) energy. One example on this is shown in Fig. A1.6
of Appendix 1.

\hspace*{3mm}
Fig.\ 5 shows an example of a reaction on a heavier target, $^9$Be, 
namely the reaction 1 GeV/nucleon $^{86}$Kr + $^{9}$Be measured
by Voss \cite{Voss95}, compared with our LAQGSM+GEM2 results.
No fission mechanism is involved in this reaction and all the
measured products published in \cite{Voss95} and shown in this 
figure are described by our code using

\end{minipage}

\end{figure}
%**************************** End Figs. 3,4  **************************

%\newpage
%**************************** Begin Fig. 5 **************************
\begin{figure}[t!]
\begin{minipage}{7.0cm}
\vbox to 9.8cm {
\vspace*{-40mm}
\hspace*{-22mm}
\includegraphics[width=100mm,angle=-0]{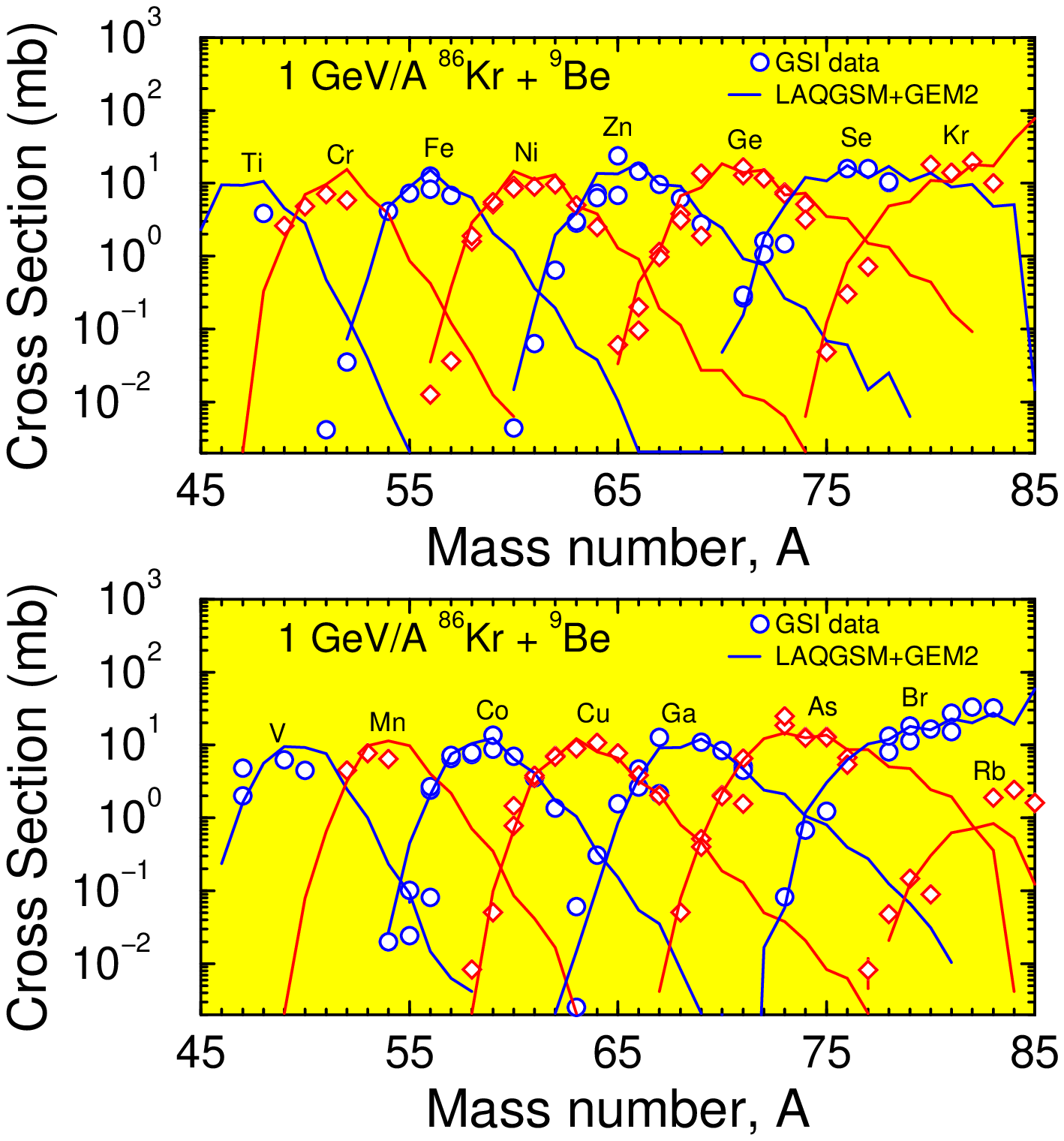}}
%\special{em:graph {e:/emtex/mytex/msp/mashnik2.msp}} }
\end{minipage}
\hfill
\begin{minipage}{10.5cm}
\vspace*{-60mm}
\begin{small}
{\bf Figure 5.}
Comparison of all measured \cite{Voss95}
cross sections of products from the reaction $^{86}$Kr + $^{9}$Be
at 1 GeV/nucleon (symbols) with our LAQGSM+GEM2 results (lines).\\
\\
\end{small}

\vspace*{-4mm}
\hspace*{-2mm}
only 
spallation. Although LAQGSM+GEM2
underestimates significantly the yields of neutron-rich Rb isotopes,
otherwise there is a good agreement between the calculations and data
for all the other measured cross sections.

\hspace*{3mm}
Fig.\ 6 shows an example of a reaction on a heavier target, $^{27}$Al, 
namely the reaction 790 MeV/nucleon  $^{129}$Xe + $^{27}$Al measured
at GSI by Reinhold {\it et al.} \cite{Reinhold98} and compared with
LAQGSM+GEM2 results. Although both the projectile and target are heavier
than for the example shown in Fig.\ 5, LAQGSM+GEM2 describes all the
products from the reaction shown in Fig.\ 6 as well using only
spallation. A very good agreement between the data and calculations
may be seen for all measured cross sections, except for the neutron-rich
Cs isotopes, whose charge is bigger than that of initial Xe nuclei of the 
beam, being produced by picking up a proton from the Al target rather than
by spallation processes. The situation observed in Fig.\ 5
for the production of neutron-rich Rb isotopes involves the same process.

\end{minipage}

\begin{minipage}{7.0cm}
\vbox to 85mm {
\vspace*{-79mm}
\hspace*{-22mm}
\includegraphics[width=100mm,angle=-0]{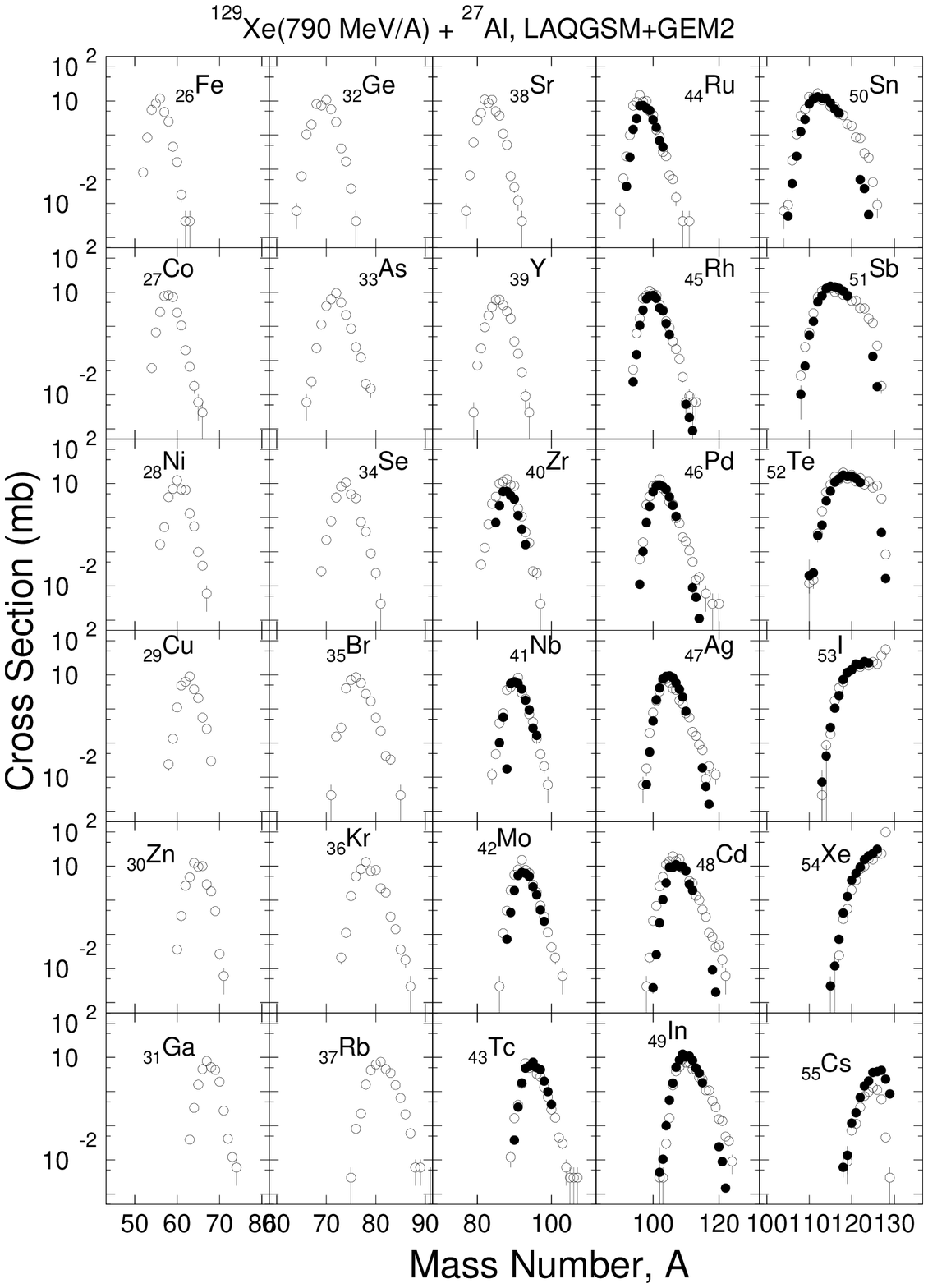}}
\end{minipage}
\hfill
\begin{minipage}{10.5cm}
\vspace*{-58mm}
\begin{small}
{\bf Figure 6.}
Comparison of all measured \cite{Reinhold98}
cross sections of products from the reaction $^{129}$Xe + $^{27}$Al
at 790 MeV/nucleon (filled circles) with our LAQGSM+GEM2 results 
(open circles). Isotopes from Fe to Y are not measured yet and we
present here our predictions.\\
\\
\end{small}

\vspace*{-4mm}
\hspace*{3mm}
Finally, Fig.\ A1.7 (Appendix 1)
shows a heavy-ion-induced reaction measured at GSI  
\cite{Junghans97,Junghans98}, namely the yields of measured spallation
products from the interaction of a 950 MeV/nucleon $^{238}$U beam 
with copper compared with our
results. LAQGSM+GEM2 describes most of these
data with an accuracy of a factor of two or better.

\hspace*{3mm}
Fig. A1.8 (Appendix 1)
shows an example of several exotic reactions, namely 
fragmentation of secondary beams of neutron-rich unstable
$^{19,20,21}$O and stable $^{17,18}$O isotopes on $^{12}$C targets
at beam energies near
600 MeV/nucleon measured recently
at GSI \cite{Leistenschneider02}, compared with our LAQGSM+GEM2 results.
The secondary beams of $^{17-21}$O ions were produced in the fragmentation
of a primary $^{40}$Ar beam at 720 MeV/nucleon on a beryllium target
(see more details in \cite{Leistenschneider02}).
A detailed discussion of our results on this reaction and more
examples of nucleus-nucleus reactions analyzed with LAQGSM+GEM2
may be found in \cite{Varenna03}.      

\end{minipage}

\vspace*{-30mm}
\hspace*{3mm}
From the results presented here and in the cited references,
we conclude that CEM2k and LAQGSM describe well (and
without any refitted parameters) a large variety of medium- and high-energy 
nuclear reactions induced both by nuclei and particles and are suitable for 
evaluations of nuclear data for applications and to study basic
problems in nuclear reaction science.
Merging our CEM2k and LAQGSM with the code GEM2 by 
Furihata \cite{GEM2} allows us to describe
reasonably well many fission and fragmentation reactions 
in addition to the spallation reactions already described 
well by CEM2k and LAQGSM. This does not mean that 
our codes are without problems.  For instance, 
LAQGSM+GEM2 does not reproduce well the mass distributions for 
some fission-fragment elements from the reaction 1 GeV/A 
$^{238}$U + $^{208}$Pb measured recently at GSI \cite{Enqvist99},
although it still reproduces very well the
integrated mass- and charge-distributions of all products.
We think that the main reasons for
this problem are the facts that the current version of LAQGSM
does not take into account electromagnetic-induced fission
\cite{Heinz03}, and because the GEM2 code by Furihata merged at
present with our LAQGSM does not consider at all the 
angular momentum of emitted particles, and of the compound nuclei.
Both these factors are especially important for reactions with heavy ions
and less important for reactions with light ions or protons;
this would explain why the code works well in the case of
reactions induced by particles and light and medium nuclei but
fails in the case of U+Pb. 
%Besides the problem of angular momentum,
%the current version of GEM2 has several more drawbacks related to
%its lack of self-consistency (see details in \cite{Mashnik02a}).
%We may choose to use a model similar to the GEM2 approach in 
%the future versions
%of our codes, but it must be significantly extended and further improved.
Our work on CEM2k and LAQGSM is not completed; we continue
their further development and improvement. Some details of our 
present work and plan
for future may be found in Refs. \cite{Varenna03,OurNewINC}.         

\vspace*{-2cm}
\end{figure}

\vspace*{-2cm}

%**************************** End Fig. 6  **************************

%\newpage
%**************************** Begin Fig. 5 **************************
\begin{figure}[t!]

This study was supported by the U.\ S.\ Department of Energy,  the
Moldovan-U.\ S.\ Bilateral Grants Program, CRDF Project 
MP2-3045, and by the NASA ATP01 Grant NRA-01-01-ATP-066.

%----- BIBLIOGRAPHY -----

\end{figure}

%\newpage

\begin{figure}[t!]

\vspace{-.5cm}
{\center \large \bf{ Appendix 1}}

\vspace{-3.5cm}
\centerline{\hspace{-8mm} \epsfxsize 20cm \epsffile{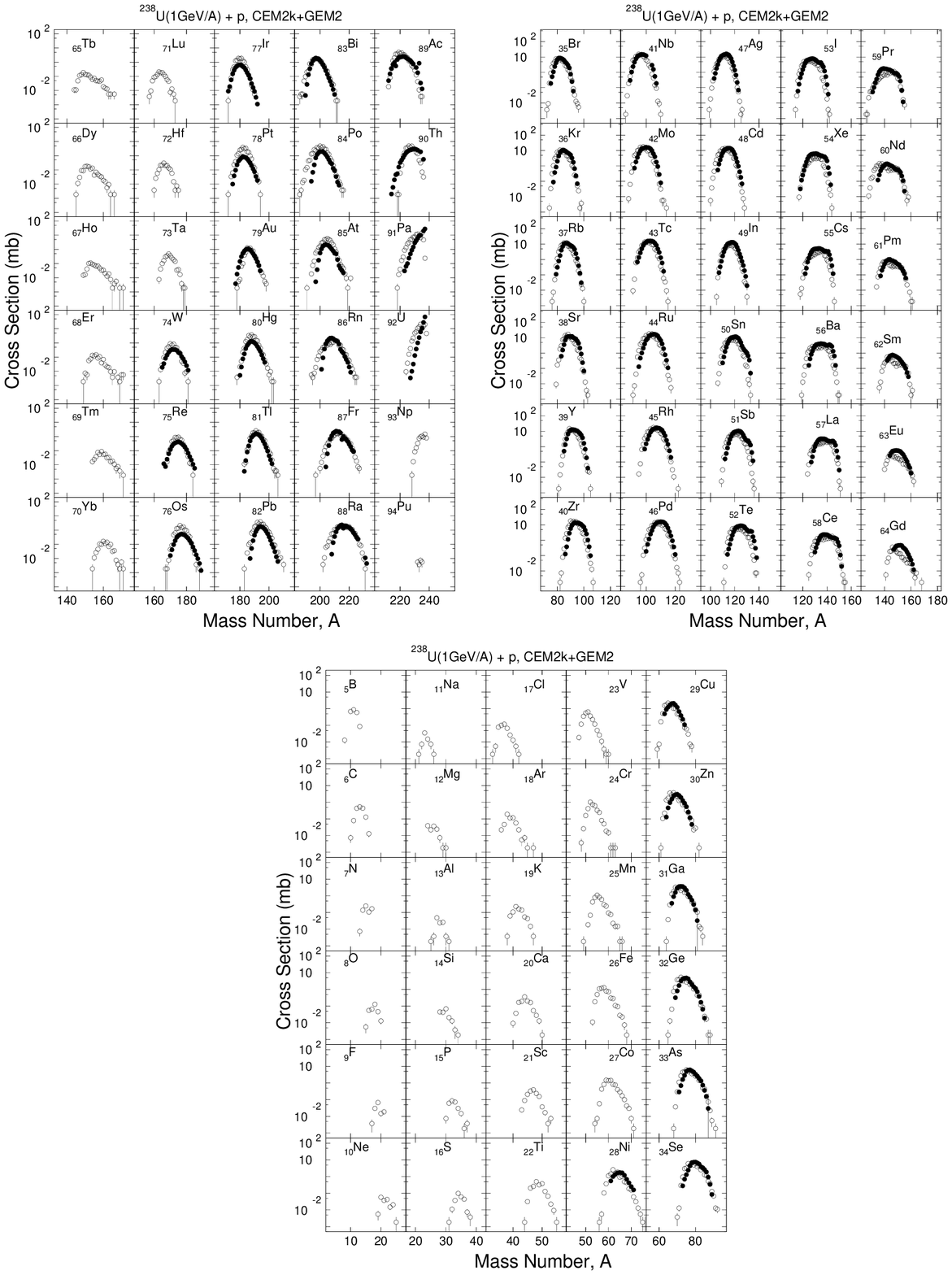}} 
\vspace{-3.0cm}
{\bf Figure A1.1.}
Comparison of measured \cite{Taieb02,Bernas03} 
spallation, fission, and fragmentation product cross
sections of the reaction $^{238}$U(1 GeV/A) + p (filled circles)
with our CEM2k+GEM2 results (open circles). Experimental data for
isotopes from B to Co, from Tb to Ta,  and for Np and Pu
are not yet available so we present here only our predictions. 
Our calculations were done in 2002 and were published  partially
in a 2002 LANL Activity Report \cite{T&NW2002}
while the fission data \cite{Bernas03} were published and become 
available to us only in 2003.
\end{figure}

%\newpage

\begin{figure}[t!]

\vspace{-3.cm}
\centerline{\hspace{-8mm} \epsfxsize 20cm \epsffile{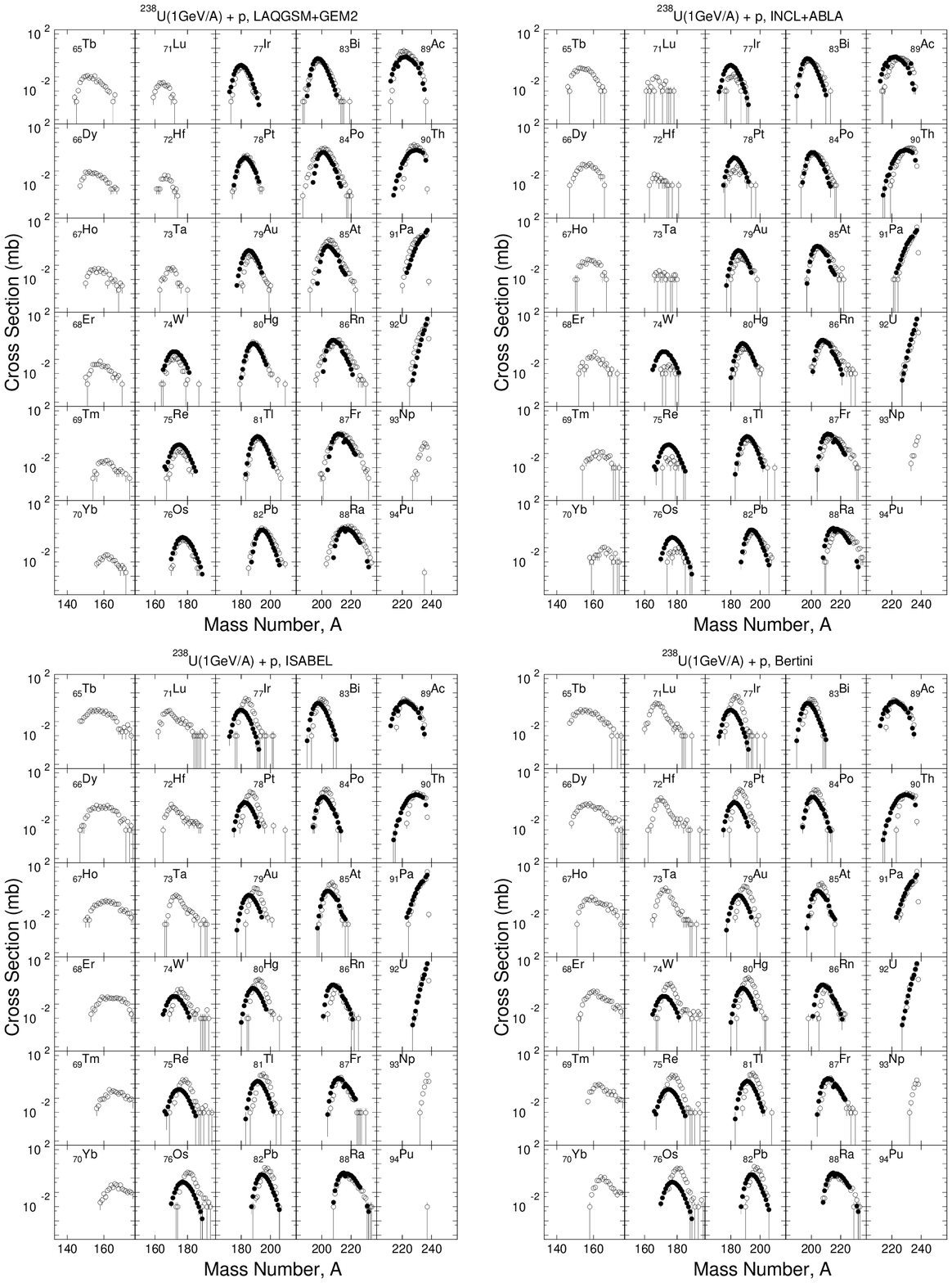}} 
\vspace{-3.0cm}
{\bf Figure A1.2.}
Comparison of measured \cite{Taieb02}  spallation product cross
sections of the reaction $^{238}$U(1 GeV/A) + p (filled circles)
with results by LAQGSM+GEM2 and by LAHET3 using the
INCL+ABLA, ISABEL+Dresner/Atchison,
and Bertini+Dresner/Atchison options (open circles), respectively. 
Experimental data for isotopes from Tb to Ta and for Np and Pu
are not yet available so we present here only our predictions.
\end{figure}

%\newpage

\begin{figure}[t!]
\vspace{-4.cm}
\centerline{\hspace{-8mm} \epsfxsize 20cm \epsffile{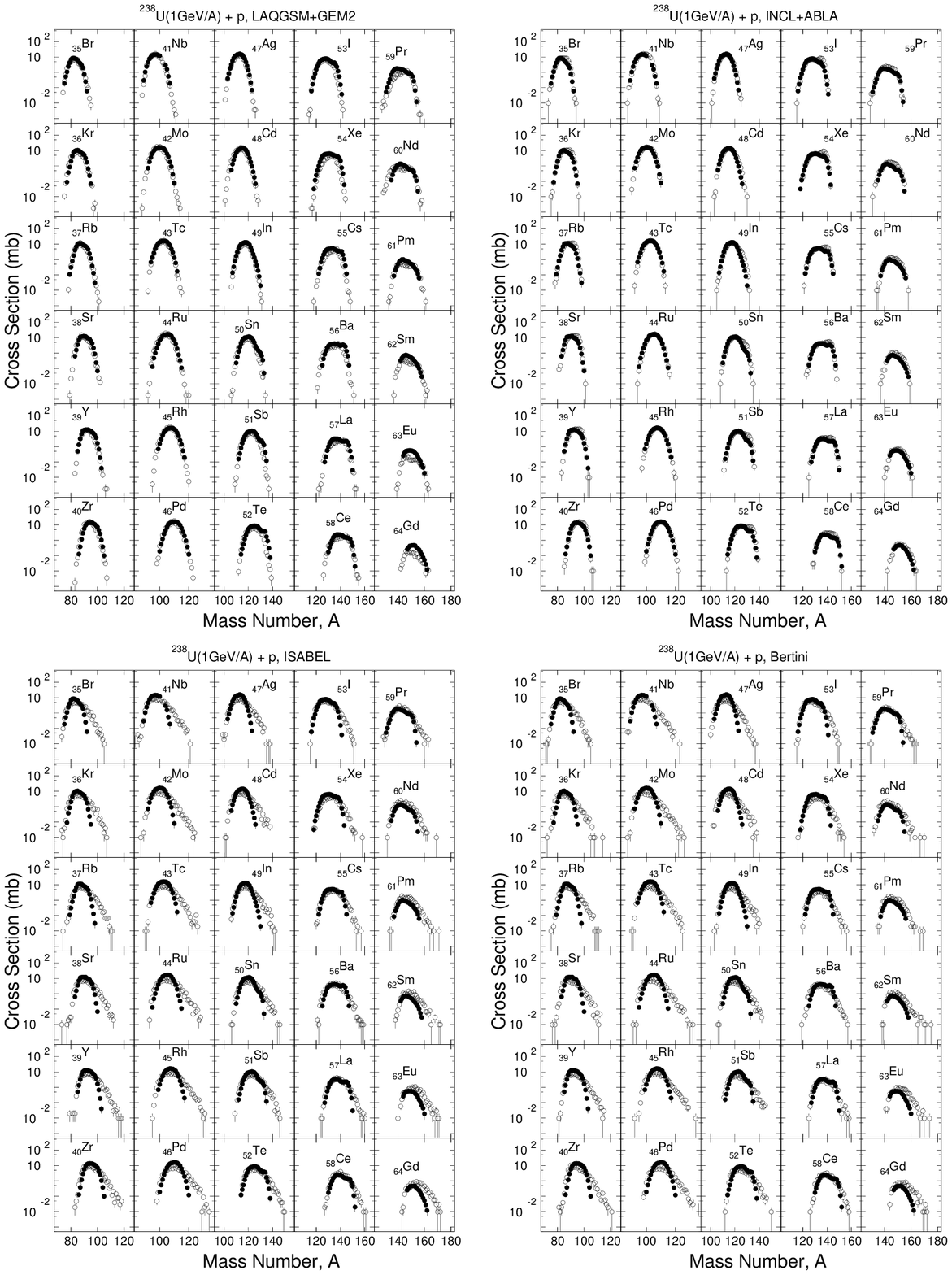}} 
\vspace{-3.0cm}
{\bf Figure A1.3.}
Comparison of measured \cite{Bernas03} fission-product cross
sections of the reaction $^{238}$U(1 GeV/A) + p (filled circles)
with results by LAQGSM+GEM2 and by LAHET3 using the
INCL+ABLA, ISABEL+Dresner/Atchison,
and Bertini+Dresner/Atchison options (open circles), respectively. 
All our calculations 
(except for the INCL+ABLA, see details in Appendix 2)
were done in 2002 and were published partially
in a 2002 LANL Activity Report \cite{T&NW2002}
while the data \cite{Bernas03} were published and become 
available to us only in 2003.
\end{figure}

%\newpage

\begin{figure}[t!]
\vspace{-4.cm}
\centerline{\hspace{-8mm} \epsfxsize 20cm \epsffile{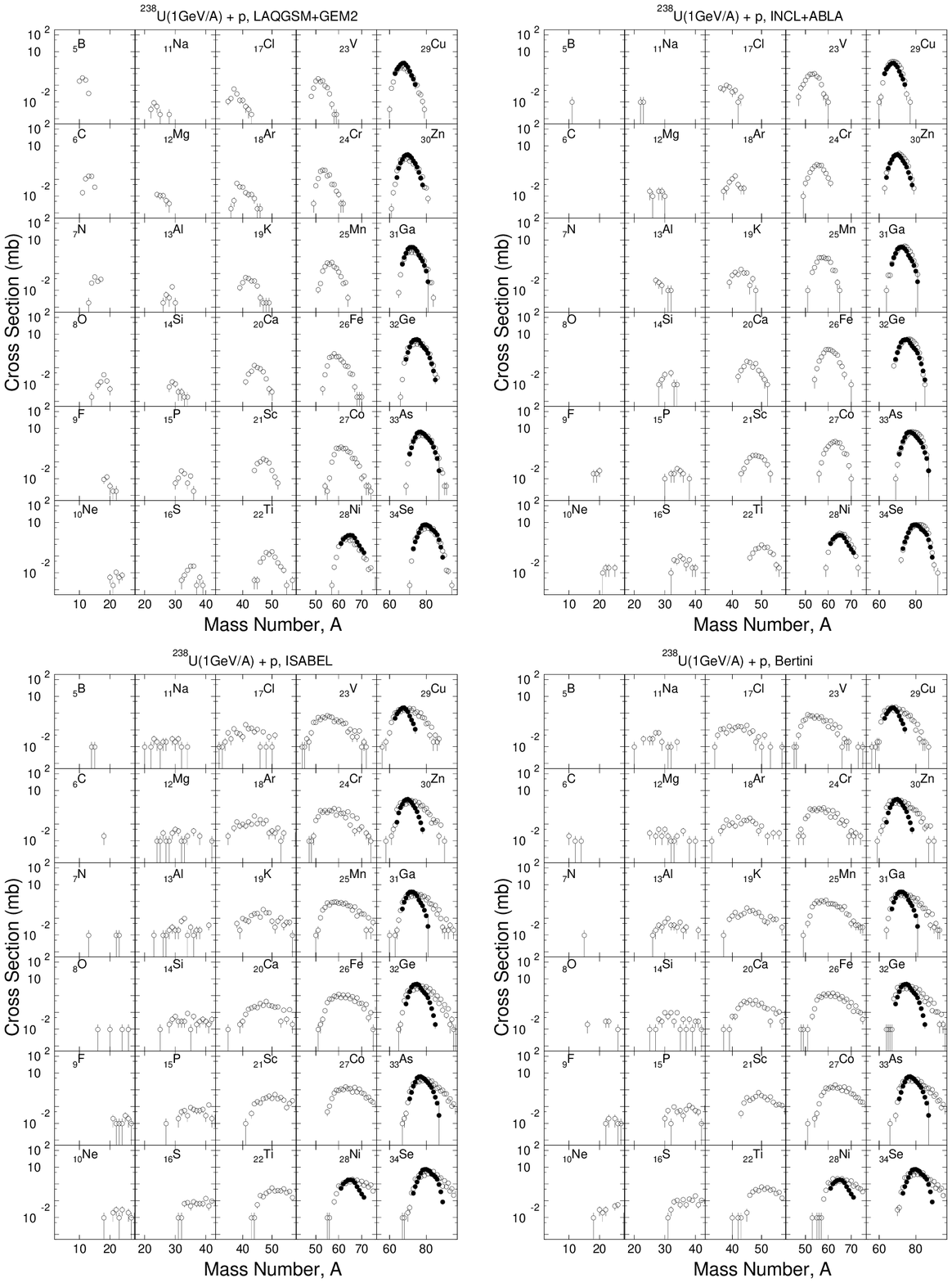}} 
\vspace{-3.0cm}
{\bf Figure A1.4.}
Comparison of measured \cite{Bernas03} fragmentation (and fission)
product cross
sections of the reaction $^{238}$U(1 GeV/A) + p (filled circles)
with results by LAQGSM+GEM2 and by LAHET3 using the
INCL+ABLA, ISABEL+Dresner/Atchison, and Bertini+Dresner/Atchison
 options (open circles), respectively.  Experimental data for isotopes  
from B to Co are not yet available
so we present here only our predictions.
All our calculations 
(except for the INCL+ABLA, see details in Appendix 2)
were done in 2002 and were published partially
in a 2002 LANL Activity Report \cite{T&NW2002}
while the data \cite{Bernas03} were published and become 
available to us only in 2003.
\end{figure}

%\newpage

\begin{figure}[t!]
\vspace{-4.cm}
\centerline{\hspace{-8mm} \epsfxsize 20cm \epsffile{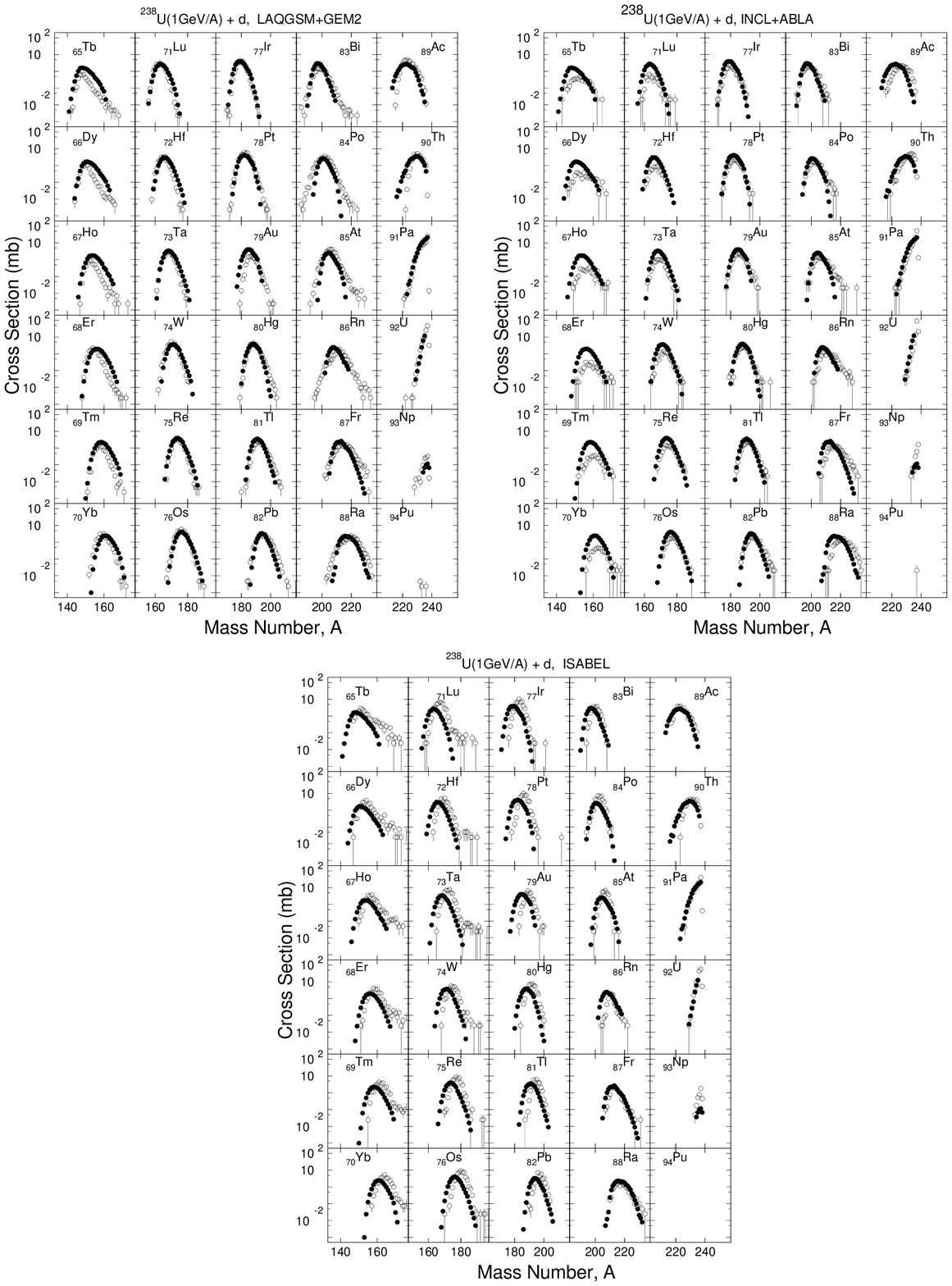}} 
\vspace{-3.0cm}
{\bf Figure A1.5.}
Comparison of measured \cite{Casarejos}  spallation-product cross
sections of the reaction $^{238}$U(1 GeV/A) + d (filled circles)
with results by LAQGSM+GEM2 and by LAHET3 using the
INCL+ABLA and ISABEL+Dresner/Atchison options
(open circles), respectively.  Experimental data for Pu isotopes
are not yet available so we present here only our predictions. 
\end{figure}	

%\newpage
\begin{figure}[h]

\vspace*{-45mm}
\hspace*{-15mm}
\includegraphics[width=190mm,angle=-0]{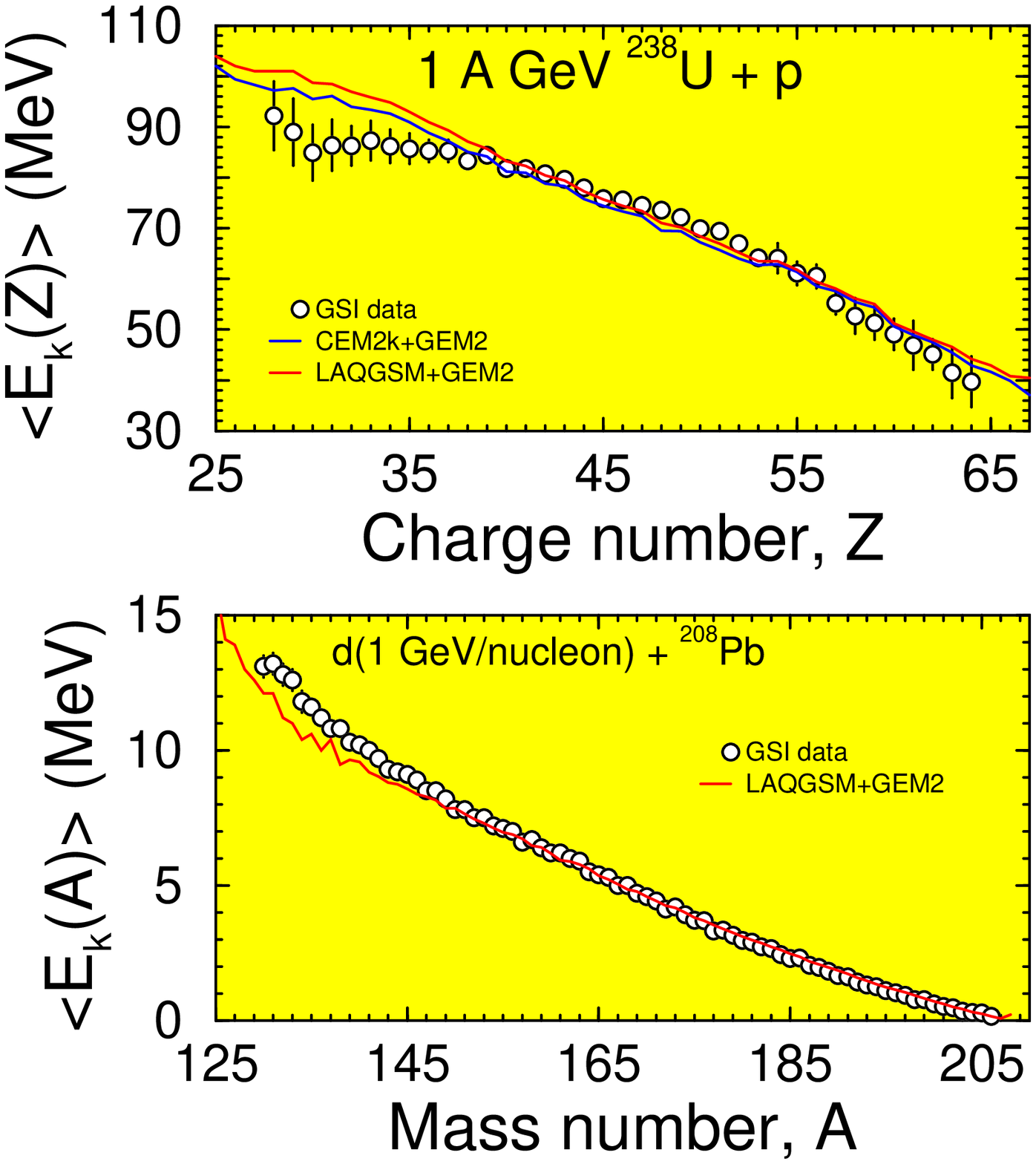}

\vspace*{-67mm}
{\bf Figure A1.6.}
Comparison of the calculated by CEM2k+GEM2 and LAQGSM+GEM2
charge distributions of the mean kinetic 
energy of nuclides produced in 1 GeV/A $^{238}$U +p and
calculated by LAQGSM+GEM2
mass distribution of the mean kinetic 
energy of nuclides produced in 
1 GeV/A $^{208}$Pb + d reactions
(lines) with the GSI measurements (circles), respectively:
The U data are from Ref. \cite{Bernas03} and the Pb data,
form \cite{Enqvist02}.\\

\vspace*{4cm}
\end{figure}

%\newpage
\begin{figure}[h]

\vspace*{-65mm}
\hspace*{-20mm}
\includegraphics[width=200mm,angle=-0]{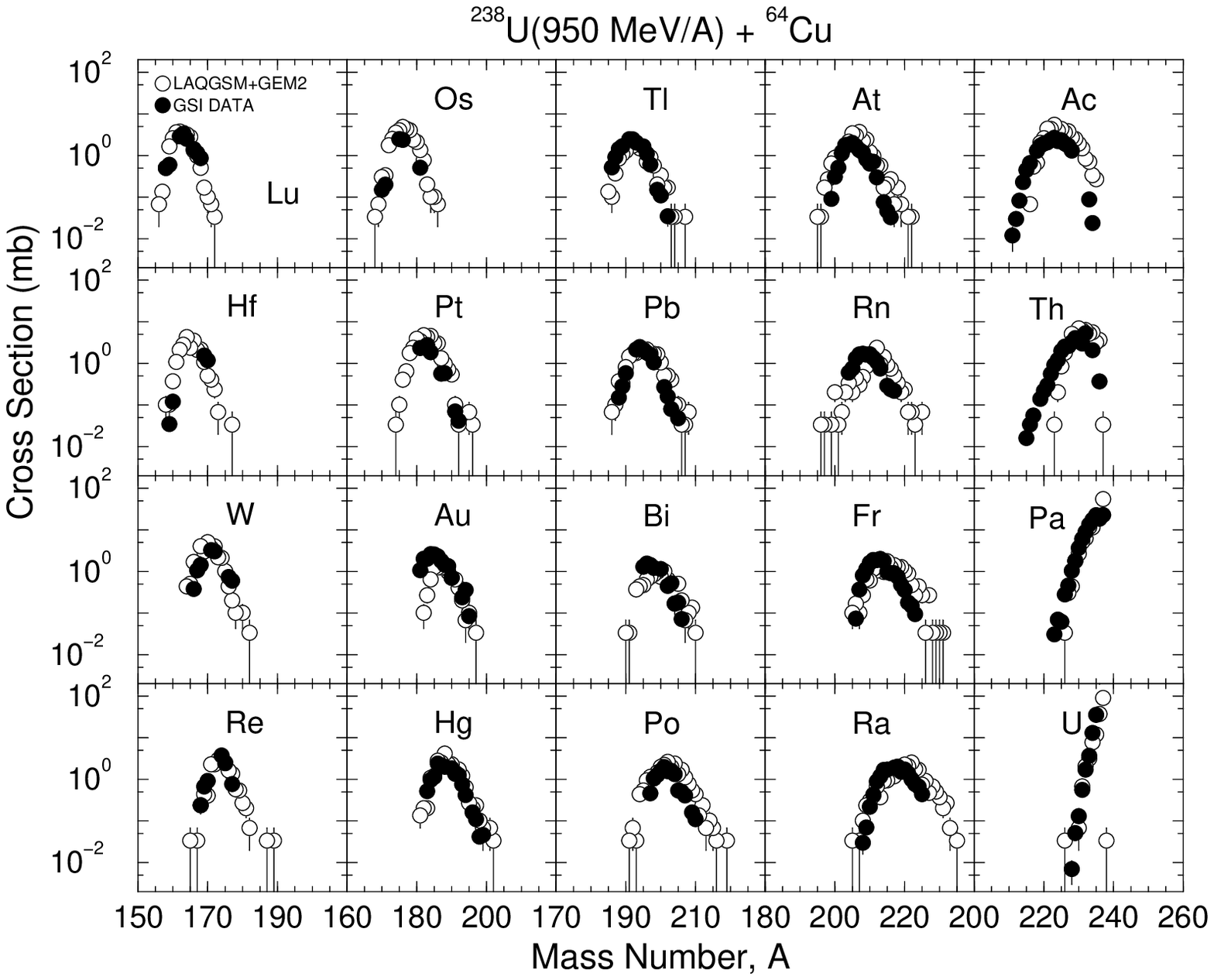}

\vspace*{-125mm}
{\bf Figure A1.7.}
Comparison of all measured \cite{Junghans97,Junghans98}
cross sections of products from the reaction $^{238}$U + $^{64}$Cu
at 950 MeV/nucleon (filled circles) with our LAQGSM+GEM2 results 
(open circles).

\end{figure}

%\newpage
\begin{figure}[h]

\vspace*{-35mm}
\hspace*{-25mm}
\includegraphics[width=200mm,angle=-0]{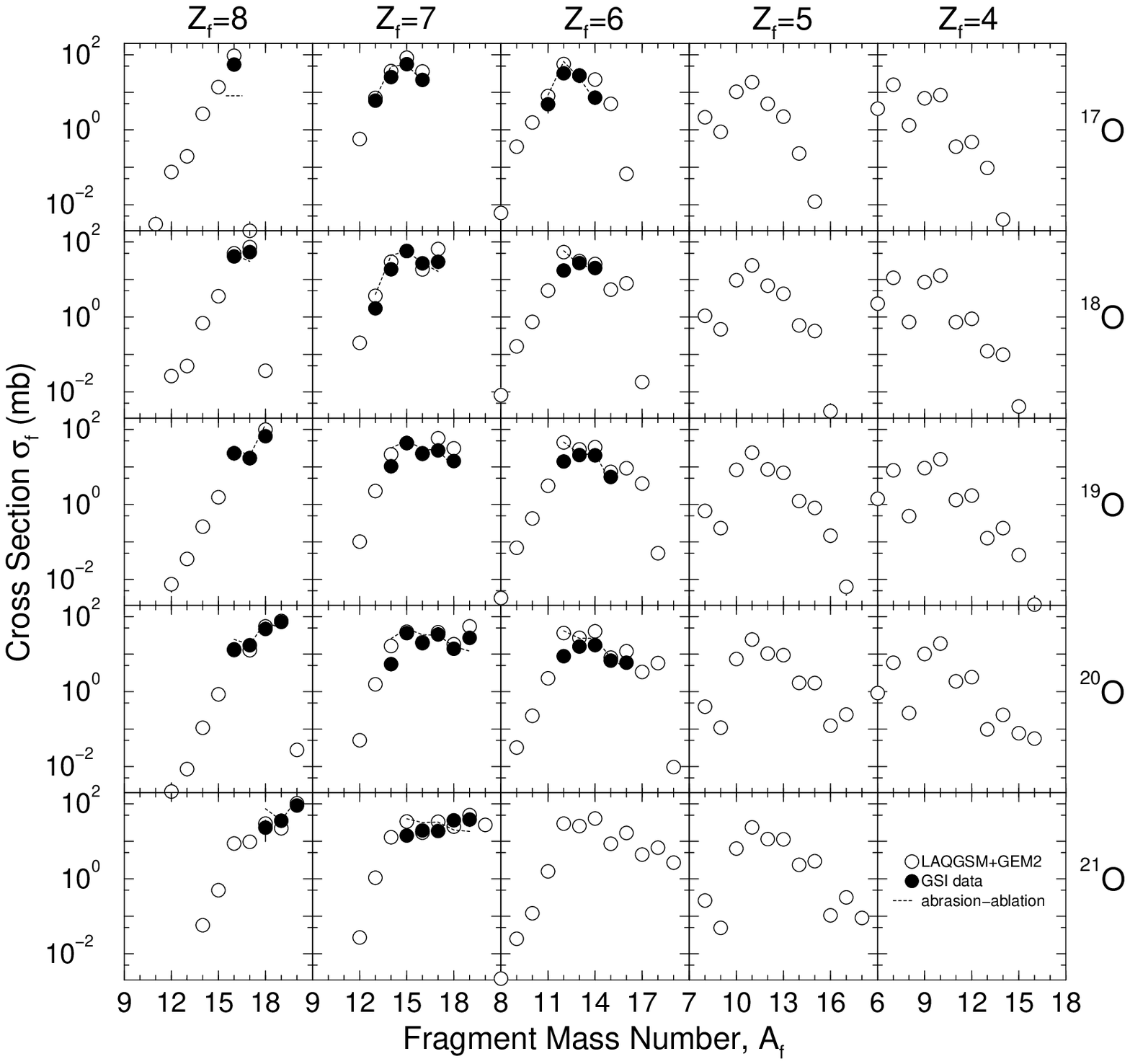}

\vspace*{-97mm}
{\bf Figure A1.8.}
Cross sections of projectile fragments with nuclear charges $Z_f$
(shown on the top) and masses $A_f$ (shown on the bottom) produced
from $^{17-21}$O beams (shown on the right) in a $^{12}$C target.
Experimental data (filled circles) are from \cite{Leistenschneider02}.
Open circles show our LAQGSM+GEM2 results for the measured cross 
sections and predictions for several unmeasured isotopes.
Dashed lines show results by the the abrasion-ablation 
model \cite{Gaimard91} from \cite{Leistenschneider02}.

\vspace*{4cm}
\end{figure}
%**************************** End Figs. A1.6&7  **************************

%\newpage

\vspace*{-3cm}
\begin{figure}[t!]

\vspace*{-7cm}

{\Large \bf Appendix 2}\\

We apologize for showing at the TRAMU workshop
results calculated with the transport code LAHET3 \cite{LAHET3}
using the Liege intranuclear cascade model INCL by Cugnon {\it et al.}
\cite{INCL} and citing them as calculated using the evaporation/fission
model ABLA by K.-H. Schmidt {\it et al.} \cite{ABLA}, i.e., INCL+ABLA.
The truth is that one of us performed these calculations with
INCL using the default options of LAHET3 for evaporation and fission, 
which selects the Dresner evaporation model \cite{Dresner}
in conjunction with the Atchison fission model 
%(often referred in the literature as RAL fission model) 
\cite{RAL} but not ABLA.
We thank Dr. Sylvie Leray to calling our attention
to this confusion. To correct ourselves and to see how using different
evaporation/fission models coupled with the same intranuclear cascade (INC)
model affects the final results
we performed after the TRAMU workshop calculations with LAHET3
using the right INCL+ABLA option. In the following three pages,
we present several figures with mass and charge distributions
of the products from all reactions we discussed at TRAMU
calculated with INCL+ABLA compared with similar results 
by INCL+Dresner/Atchison
showed at TRAMU (and cited mistakenly as INCL+ABLA) and by 
Bertini+Dresner/Atchison, ISABEL+Dresner/Atchison, 
CEM2k+GEM2, and LAQGSM+GEM2 models discussed in our talk
at TRAMU. One may see a big difference between INCL+ABLA
and INCL+Dresner/Atchison results.  On the whole,
INCL+ABLA provides a better
agreement with the measured fission products than when using the
Dresner/Atchison option to calculate evaporation and fission,
but strongly underestimates the spallation products, especially with
masses near the border between the spallation and fission regions,
and agrees much worse with the data in this mass region than
INCL+Dresner/Atchison does for these products.
We see that
different evaporation/fission models coupled with the same INC model
provide significantly different results in both the
spallation and fission regions of products, suggesting 
that development of a reliable and universal evaporation/fission model
is of a first priority for any transport code, independently of what
INC model is used to describe the intranuclear-cascade stage of a reaction.

Finally, we would like to warn future users of LAHET3 that when
choosing the INCL option for the intranuclear cascade model in LAHET3,
the default options for evaporation will be Dresner and for fission,
Atchison, and one needs to specify explicitely the option ABLA to
get results by INCL+ABLA, otherwise LAHET3 will provide results
by INCL+Dresner/Atchison, as happened to us before TRAMU.

\end{figure}

%\newpage

\begin{figure}[t!]
\vspace{-5.cm}
\centerline{\hspace{-8mm} \epsfxsize 20cm \epsffile{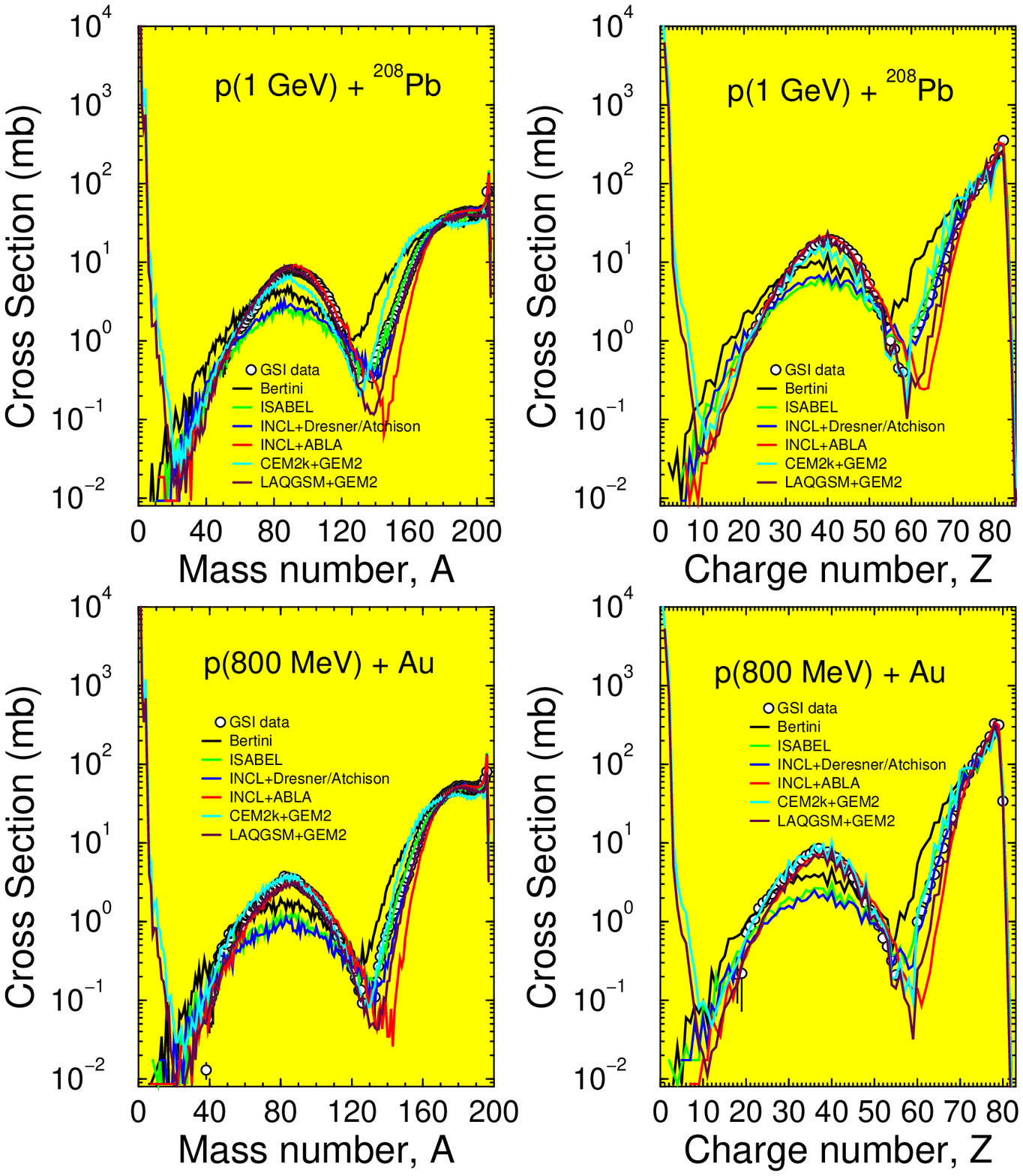}} 
\vspace{-4.0cm}
{\bf Figure A2.1.}
Comparison of measured 
mass (left panel) and charge (right panel) distributions
of the nuclides produced in the reactions 1 GeV/A $^{208}$Pb + p
and 800 MeV/A $^{197}$Au + p 
with results by LAHET3 using the Bertini+Dresner/Atchison, 
ISABEL+Dresner/Atchison,
INCL+Dresner/Atchison, and the INCL+ABLA options, and by the CEM2k+GEM2 and
LAQGSM+GEM2 codes
(color lines), respectively. 
Experimental data for $^{208}$Pb
are from Ref. \cite{Enqvist01} and for $^{197}$Au, from
Ref. \cite{Rejmund01}.
\end{figure}	

%\newpage

\begin{figure}[t!]
\vspace{-5.cm}
\centerline{\hspace{-8mm} \epsfxsize 20cm \epsffile{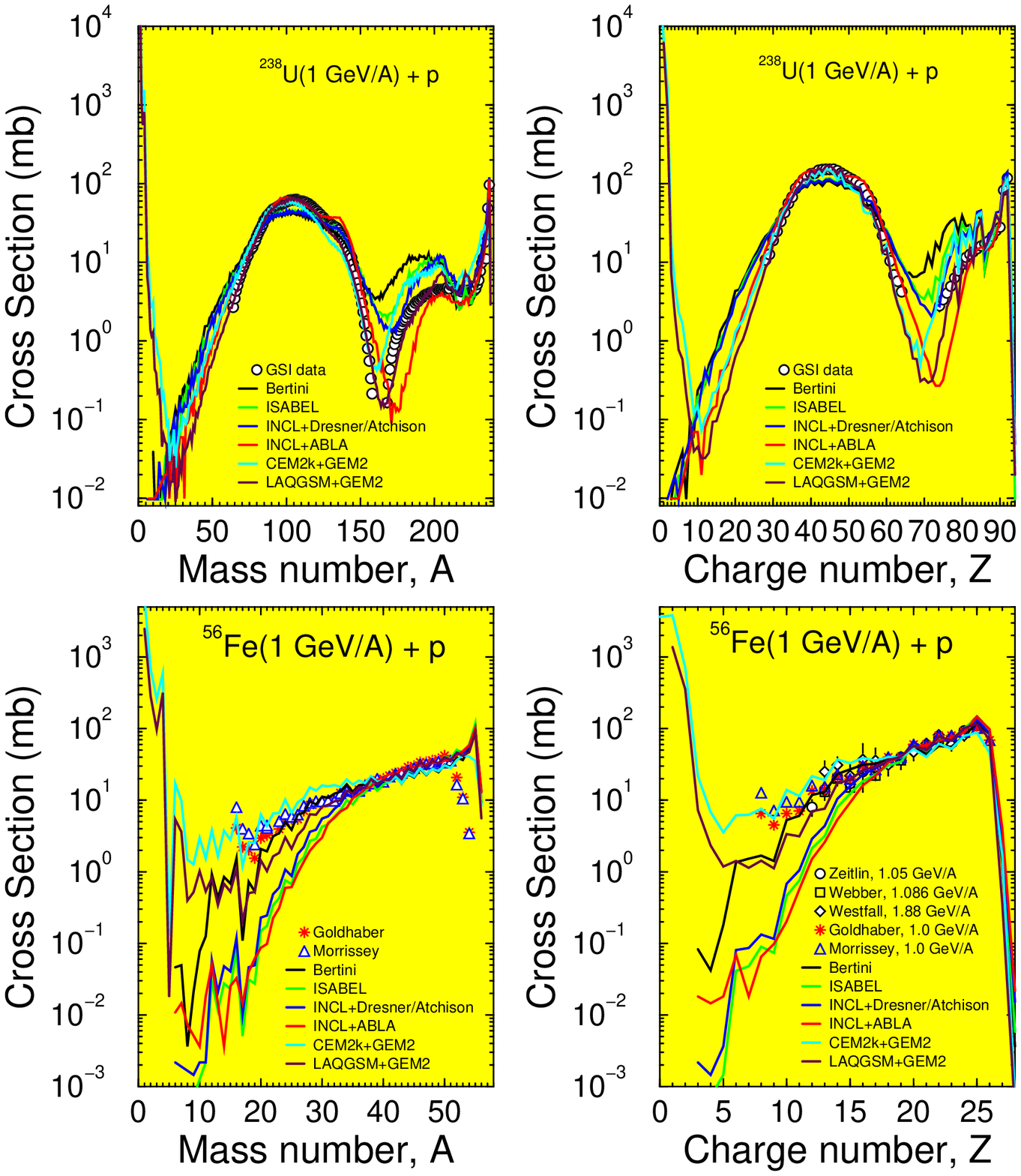}} 
\vspace{-4.0cm}
{\bf Figure A2.2.}
Comparison of measured mass (left panel) and charge (right panel) 
distributions of the nuclides produced in the reactions 1 
GeV/A $^{238}$U + p and 1 GeV/A $^{56}$Fe + p with results from 
LAHET3 using the Bertini+Dresner/Atchison,  
ISABEL+Dresner/Atchison,
INCL+Dresner/Atchison, and the INCL+ABLA options,
and by the CEM2k+GEM2 and LAQGSM+GEM2 codes
(color lines), respectively. 
Experimental data for $^{238}$U are from Refs. \cite{Taieb02,Bernas03}.
The 1 GeV/A GSI data for $^{56}$Fe were obtained
using the systematics by Morrissey \cite{Morrissey} and the
Goldhaber model \cite{Goldhaber} 
and are extracted by us from Figs. 7 and 8 of Carmen Villagrassa
\cite{Carmen}; earlier Fe data at 1.05 GeV/A by Zeitlin {\it et al.} \
\cite{Zeitlin}, 1.086 GeV/A by  Webber {\it et al}. \cite{Webber},
and 1.88 GeV/A by  Westfall {\it et al.} \cite{Westfall}
(all tabulated in Tab. 7 of Ref. \cite{Zeitlin}) are also
shown for comparison.

\end{figure}

%\newpage

\begin{figure}[t!]
\vspace{-5.cm}
\centerline{\hspace{-8mm} \epsfxsize 20cm \epsffile{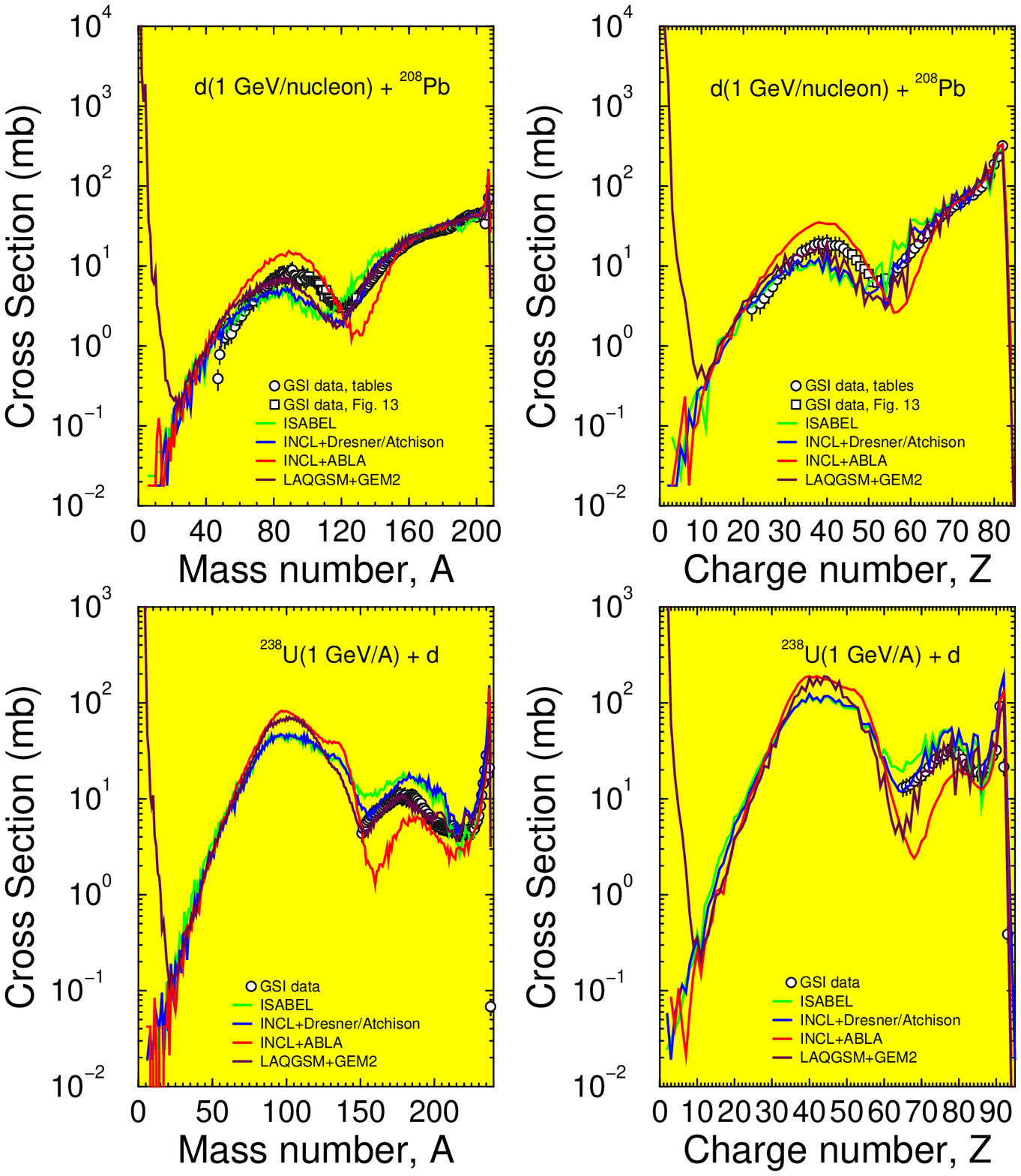}} 
\vspace{-4.0cm}
{\bf Figure A2.3.}
Comparison of measured 
mass (left panel) and charge (right panel) distributions
of the nuclides produced in the reactions 1 GeV/A $^{208}$Pb + d
and 1 GeV/A $^{238}$U + d  with results from LAHET3 using the
ISABEL+Dresner/Atchison,
INCL+Dresner/Atchison, and the INCL+ABLA options, and by our
LAQGSM+GEM2 
(color lines), respectively. 
Experimental data for $^{208}$Pb
are from Tabs. 2 and 3 and Fig. 13 of Ref. \cite{Enqvist02} 
and for $^{238}$U, from Ref. \cite{Casarejos}.
\end{figure}

\end{document}